\documentclass[sigconf]{acmart}

\usepackage[table]{xcolor}  
\usepackage{enumitem}       
\usepackage{multirow}
\usepackage{rotating}
\usepackage{tabularx}
\usepackage{longtable}
\usepackage{placeins}

\definecolor{orange}{rgb}{1.0, 0.5, 0.0}
\definecolor{purple}{rgb}{1.0, 0.0, 0.5}
\definecolor{darkgreen}{rgb}{0.0, 0.8, 0.0}
\definecolor{lightblue}{rgb}{0.6, 0.7, 1.0}

\copyrightyear{2026}
\acmYear{2026}
\setcopyright{cc}
\setcctype{by}
\acmConference[ASSETS '26]{The 28th International ACM SIGACCESS Conference on Computers and Accessibility}{October 25--28, 2026}{Vila Nova de Gaia, Portugal}
\acmBooktitle{The 28th International ACM SIGACCESS Conference on Computers and Accessibility (ASSETS '26), October 25--28, 2026, Vila Nova de Gaia, Portugal}
\acmDOI{10.1145/3797867.3829033}
\acmISBN{979-8-4007-2521-0/2026/10}

\begin{document}

\raggedbottom

\title{Leveling the Playing Field: Temporal Video Segmentation for Individuals with ADHD in Computing Education}

\author{Veronica Pimenova}
\affiliation{%
 \institution{University of Michigan}
 \city{Ann Arbor}
 \state{Michigan}
 \country{USA}}
\email{pimenova@umich.edu}

\author{Chris Lee}
\affiliation{%
  \institution{Carnegie Mellon University}
  \city{Pittsburgh}
  \state{Pennsylvania}
  \country{USA}}
\email{cylee@andrew.cmu.edu}

\author{Baramee Bhakdibhumi}
\affiliation{%
  \institution{Carnegie Mellon University}
  \city{Pittsburgh}
  \state{Pennsylvania}
  \country{USA}}
\email{barameeb@andrew.cmu.edu}

\author{Simon Chu}
\affiliation{%
  \institution{Carnegie Mellon University}
  \city{Pittsburgh}
  \state{Pennsylvania}
  \country{USA}}
\email{cchu.simon@gmail.com}

\author{Andrew Begel}
\affiliation{%
  \institution{Carnegie Mellon University}
  \city{Pittsburgh}
  \state{Pennsylvania}
  \country{USA}}
\email{abegel@cmu.edu}

\renewcommand{\shortauthors}{Pimenova et al.}

\begin{abstract}
Individuals with Attention-Deficit/Hyperactivity Disorder (ADHD) often face significant barriers in computing education. In asynchronous learning environments, instructional videos can impose high \emph{extraneous} cognitive load, often relying on assumptions about sustained attention and working memory that do not align with ADHD neurocognitive profiles. In this work, we evaluate a post-hoc video processing intervention that segments instructional content into single-instruction chunks followed by fixed-length pauses to reduce cognitive load. In a within-participants controlled study with 17 individuals with ADHD and 10 without, we find that the intervention has an equalizing effect. Although it improved performance for all participants, gains were larger for those with ADHD, reducing their errors and hesitations to levels comparable to those of participants without ADHD under the same intervention. These results align with the goals of Universal Design for Learning (UDL), by showing that cognitively-aligned, post-hoc instructional video modifications can reduce performance disparities across diverse neurocognitive profiles. 
\end{abstract}



\noindent

\noindent

\noindent

\noindent

\ccsdesc[500]{Human-centered computing~Accessibility systems and tools}
\ccsdesc[300]{Human-centered computing~Empirical studies in HCI}
\ccsdesc[300]{Applied computing~Interactive learning environments}
\ccsdesc[300]{Software and its engineering~Software usability}
\ccsdesc[300]{Social and professional topics~Assistive technologies}
\ccsdesc[300]{Computing methodologies~Learning paradigms}

\keywords{ADHD, Cognitive Load, Accessibility, Programming Education}

\maketitle

\section{Introduction}

\textit{Attention-Deficit Hyperactivity Disorder} (ADHD) 
is a cognitive condition characterized by differences in working memory, the cognitive system that supports the temporary storage and manipulation of information~\cite{baddeley1992working, apa2025adhd}. Working memory performance is especially salient for cognitive tasks which require problem solving or learning. This can present unique barriers in computing education, where traditional pedagogical materials for programming instruction require an ability to manage multiple cognitive demands with the integration of new, often complex, concepts~\cite{morrison-icer14}. With roughly 10.57\% of developers globally identifying with ADHD~\cite{stackoverflow2022, newman2025groove}, the creation and distribution of inclusive pedagogical material to teach programming has become increasingly more important, especially for individuals with ADHD.

Spurred by COVID-19, programming educational materials have increasingly added asynchronous or completely online instruction to traditional in-person instructional methods. Instructional videos have become an increasingly prevalent medium for self-directed learning, particularly on platforms such as YouTube~\cite{youtube}, Coursera~\cite{coursera2025whatis}, Udacity~\cite{udacity2026catalog}, or edX~\cite{edx2026about}. These videos enable students to learn concepts whenever they like, making computing education more accessible for a broader range of individuals.

Video-based instruction often assumes a neuro-normative baseline for cognitive processing speed and working memory. In many educational contexts, these videos are structured as continuous, high-density streams of information, often following a "rapid-fire" delivery style that prioritizes content coverage over processing time~\cite{guo2014video}. For students with ADHD, this design creates an environmental barrier, imposing a disproportionately high \textit{extraneous} cognitive load~\cite{morrison-icer14,sweller2011cognitive} (i.e., cognitive load caused by elements of instructional design that unnecessarily complicate learning) by presenting new information in a continuous flow as well as by forcing the individual to track both visual and verbal cues at once~\cite{martinelli2017investigating}. This makes it more difficult for them to sustain focus and manage their cognitive resources~\cite{shaw2022impact}. For example, an ADHD student with slower cognitive processing than an instructional video was designed for may be unable to fully comprehend a teacher's instructions at the pace that the teacher speaks, leading them to fall behind and lose track of the lesson~\cite{watkins-2024, thorlakson2010experiences}. This could happen in informal and remote learning contexts, where the absence of a live instructor removes the critical feedback loops used to ``read the room'' and adjust pacing dynamically~\cite{begel2019outside}. Without an instructor to proactively identify missing skills or provide personalized feedback, students with ADHD could be forced into a haphazard and difficult learning process where they need to self-regulate their own cognitive load~\cite{begel2019outside, begel2008novice}. 

Instead of blaming the learning problem entirely on the student with ADHD due to ``cognitive deficiencies,'' we follow the social model of disability~\cite{mankoff-assets10, brisenden1986independent}, which reframes the problem as one of environmental and systemic conditions that impose unreasonable barriers to people with disabilities ~\cite{borsotti-cscw24}. When instructional materials fail to account for varying cognitive learning styles, students are often forced to perform ``invisible labor''  as a form of self-accommodation  to bridge the gap (such as manually pausing and rewinding videos to manage information density)~\cite{ezeamii2025navigating, 2025longitudinal}. ADHD medications can assist students with attention regulation and executive function; however, they merely alter the barriers to learning, which are far from eliminated~\cite{holland2019relative}.


To make progress towards educational equity, we need to develop \textit{non-stigmatizing pedagogical strategies}, such as those that prioritize Universal Design for Learning (UDL)~\cite{rose2000universal}. Such strategies do not merely assist a specific subgroup; rather, they aim to equalize performance across diverse neurocognitive profiles by removing design-induced barriers. For example, strategies that promote flexibility, structure learning experiences, and deliver content in a more accessible way using techniques from UDL might include highlighting important concepts, timed check-ins for users, or video players that monitor learner performance during instructional tasks. 

Rather than evaluating interventions against a neuro-normative baseline, this work interrogates how digital learning environments can dynamically adapt to varying cognitive characteristics. Consequently, this study asks \textbf{what post-hoc changes can be made to a learn-by-doing video to improve  incremental information processing, support restoration of focus after distractions, and reduce cognitive load, without requiring instructors to specifically tailor their lessons for learners with ADHD?}

In our work, we propose a segmented learning intervention designed to support individuals with ADHD. Cognitive load theory and working memory theories predict that segmenting lectured information into smaller units will improve comprehension and retention~\cite{merkt2018pauses, ljubojevic2025improving,
morrison-icer14, baddeley2000episodic}. After an instructional video has been recorded, we identify where instructional content can be broken down into discrete, more digestible steps, and then inject brief pauses after each step. Our approach allows students to process information incrementally, refocus when distracted, and reduce cognitive overload. Importantly, the entire modification can be implemented in post-production, freeing the instructor from having to tailor their lesson or instructional delivery just for learners with ADHD. 

We study the impact of segmented video instruction by comparing it to unsegmented (i.e., continuous) video formats in the context of introductory programming education. We conducted a within-subjects experiment with 27 adult non-programmers, 17 of whom were diagnosed with ADHD. Participants engaged in learning-to-program tasks after watching segmented or unsegmented instructional videos. We chose to have developers program in Scratch because of its global usage and relevance in teaching beginners to code~\cite{resnick2009scratch}. Scratch has had over 135 million users and 164 million projects worldwide since its creation in 2008~\cite{scratchstats2026}. While Scratch instruction is typically targeted at primary school students, it is often used in university instruction of non-computing majors whose goal is to build computer literacy~\cite{malanscratch2007}.  

We hypothesized that segmentation would act as an equalizing support, reducing the performance gap between learners with and without ADHD. Our results confirmed this, finding that participants with ADHD exhibited an 87\% reduction in errors and a 79\% reduction in hesitations, effectively reducing their performance disparity with that of their non-ADHD peers under the same intervention. Both groups achieved performance gains, however participants with ADHD derived a substantially greater benefit from the segmentation intervention. Our findings point the way towards the potential for adaptive instructional strategies that help all learners, with ADHD and without, learn to program with equivalent success. We discuss the broader implications of our findings, and then suggest additional strategies to employ in the future. 
By recognizing the diverse cognitive strengths and learning styles of individuals with ADHD, this research contributes pedagogical interventions that could broaden participation in computing. 

This study makes the following contributions:

\begin{enumerate}
    \item The \textbf{introduction of a novel, lightweight, segmented video instruction intervention} designed to support learners with ADHD in programming education. 
    \item An \textbf{experimental investigation} and \textbf{quantitative analysis} of the effectiveness of this segmented video instruction approach, evaluating its impact on learning outcomes for both individuals with and without ADHD.
    \item \textbf{Practical insights and recommendations} for designing accessible educational tools, demonstrating how digital learning platforms can be adapted to support individuals with ADHD.
\end{enumerate}

\section{Background \& Related Work}

Our work builds on a foundation of \textit{cognitive theories related to ADHD, remote learning, remote learning for individuals with ADHD, and interventions}. We place a particular focus on computer science education through block-based programming, extending the literature by empirically evaluating the application of a segmentation intervention in programming education for individuals with ADHD. 

\subsection{Cognitive Theories Related to ADHD}
To develop effective instructional strategies for learners with ADHD, we draw on established cognitive theories that explain how individuals process and retain instructional material.

\subsubsection{Cognitive Load Theory (CLT)}
Cognitive Load Theory (CLT) explains how cognitive resources are allocated during learning. It categorizes cognitive load into three types: \textbf{intrinsic}, the inherent complexity of the material; \textbf{extraneous}, unnecessary cognitive effort caused by poor design; and \textbf{germane}, the mental effort dedicated to meaningful learning~\cite{sweller2011cognitive, orru2019evolution}. CLT emphasizes that reducing extraneous cognitive load while optimizing intrinsic and germane loads should enhance comprehension~\cite{sweller2011cognitive, brunken2010current, van2012cognitive}. Poor instructional design can disproportionately impact individuals with ADHD, hindering retention and participation in STEM fields~\cite{dupaul2009adhd, morrison-icer14}. We hypothesize that closely spaced instructor commands can overwhelm learners with ADHD, who may not have sufficient time to process one instruction before the next begins. Introducing pauses after each command should reduce extraneous cognitive load in line with CLT principles and enable more effective processing.

\subsubsection{Working Memory}
Working memory is the cognitive system responsible for the temporary storage and real-time processing of information~\cite{baddeley1992working}. It plays a central role in problem solving, comprehension, and learning, yet its limited capacity constrains cognitive load and shapes learning outcomes.
A component of working memory is the \textbf{phonological loop}, which supports temporary storage and rehearsal of verbal and auditory information~\cite{baddeley2000episodic}. It consists of the \textbf{phonological store}, which briefly retains verbal input, and the \textbf{articulatory rehearsal process}, which maintains information through repetition~\cite{baddeley1992working, baddeley2000episodic}. All of these components are essential to process verbal tasks, such as following instructions in programming education videos. 

Among learners with ADHD, differences in rehearsal process performance may result in verbal instructional information being evicted from the phonological store before the learner can act on it, thereby hindering successful task completion. Pinelli and Cojean provided empirical evidence for a ``working memory reversal effect,'' demonstrating that imposed pauses in instructional videos are significantly beneficial for learners with low working memory capacity (WMC), while offering no such benefit to high-WMC learners~\cite{pinelli2025working}. This shows that temporal modifications such as segmentation could act as a critical compensatory mechanism for the specific working memory constraints frequently observed in individuals with ADHD. Although stimulant medications mitigate some of these differences by strengthening the functional connectivity of frontoparietal networks during working memory tasks~\cite{wong2011effects}, pharmacological support alone does not prevent working memory overload.

\subsubsection{Dual-Channel Processing}
Mayer’s Cognitive Theory of Multimedia Learning (CTML) posits that humans process information through two distinct channels: one for auditory/verbal input and one for visual/pictorial input~\cite{mayer2003nine,moreno2001designing}. Instructional videos naturally engage both channels by combining spoken explanations with visual elements~\cite{vass-2011}. Designing materials that distribute cognitive load across these channels can improve comprehension and retention~\cite{vass-2011}. However, individuals with ADHD often find processing two channels of information simultaneously can result in increased cognitive load when engaging in dense or rapidly presented content~\cite{martinussen2005meta, kofler2010adhd, holmes2014}. Stimulant treatment has been shown to improve performance during selective visual and auditory attention tasks~\cite{wong2011effects, jonkman1997effects}, but it does not resolve the structural pacing challenges inherent in rapid, dual-channel media. Segmenting instructional videos with pauses supports dual-channel processing by allowing individuals the time to synchronize and integrate the information received through these two channels, promoting more effective student engagement with learning materials~\cite{mayer2003nine, merkt2018pauses}.

\subsection{Remote Learning \& Universal Design for Learning}

Since the COVID-19 pandemic, online learning has become increasingly prevalent, allowing learners to have access to instructional content outside of typical classroom settings~\cite{dhawan2020online, means2013meta}. This includes asynchronous video lessons, where students can watch pre-recorded instructional material on their own time. Self-paced video instruction has a far reach, as evidenced by platforms like Khan Academy, which has more than 120 million annual learners~\cite{khanacademy2025}.

Remote learning may also introduce distinct cognitive challenges, however, because students need to manage their attention, pacing, and mitigation of potential distractions when learning outside the classroom~\cite{moster-cseet22}. Video-based instruction can be particularly demanding, as learners need to process new information quickly, while also integrating visual and audio processing without immediate feedback. This could lead to increased misconceptions, errors, and reduced comprehension~\cite{bastian2025misconceptions}.

A method for developing inclusive instructional material is by the utilization of Universal Design for Learning (UDL)~\cite{israel2020teaching}, where instructors design pedagogical material for a range of cognitive needs. This includes multiple means of representation, action, and engagement~\cite{israel2020teaching, ross2019supporting, cast2024udl}. UDL helps students of varying cognitive needs maintain focus and content comprehension, especially in remote learning contexts~\cite{rao2021inclusive, mayer2014multimedia}. Without a live instructor available to clarify doubts or troubleshoot errors~\cite{shah2025active, manesh-icer2025}, students may feel isolated and struggle to progress when encountering technical difficulties or conceptual roadblocks~\cite{borsotti-cscw24}. Maintaining focus and motivation in a remote setting can also be difficult~\cite{le_cunff2022supporting}. The flexibility of online learning environments often comes at the cost of increased distractions, both digital and physical~\cite{das-cscw21}.

\subsection{Computing Education for ADHD}

Computing education research for neurodivergent learners, particularly those with ADHD, remains preliminary and underexplored ~\cite{zastudil2025neurodiversity, hirst2025prevalence}. Apart from Lim et al.'s evaluation of an AI-driven programming tutor for middle school students with ADHD ~\cite{lim_pypal}, Zastudil et al.'s systematic literature review found that many existing curricular recommendations lack supporting empirical evidence ~\cite{zastudil2025neurodiversity}. This is particularly problematic because STEM potential is often described through a medical model of disability that views cognitive disability as an inherent, natural barrier to technical achievement~\cite{shifrer2021problematizing, marks1997models}. Shifrer and Freeman show that while students with medicated ADHD often demonstrate high levels of STEM achievement and a higher likelihood of pursuing STEM education than their neurotypical peers, their success is frequently hindered by structural barriers rather than a lack of interest or aptitude~\cite{shifrer2021problematizing}. This follows the social model of disability~\cite{mankoff-assets10, brisenden1986independent}, where the focus shifts towards societal and environmental barriers rather than individual ones. 

These barriers are especially pronounced in post-secondary environments, where the shift toward less structured, predominantly verbal instruction (such as long lectures and rapid-fire auditory commands) can ``bump'' critical information out of a student's working memory before it can be processed or encoded~\cite{fleming2012developmental}. Nilholm explains that challenges faced by students with ADHD should be viewed as an ``educational challenge'' that requires pedagogical solutions developed within the classroom context itself, rather than external medical interventions~\cite{nilholm2014adhd_commentary}. Within computing education particularly, educational challenges are compounded by high ``intrinsic'' cognitive load of programming, where the simultaneous demands of syntax acquisition and logical problem-solving can overwhelm ADHD students' executive functioning capacity~\cite{baddeley1992working, apa2025adhd}.

\subsection{Remote Learning for ADHD}

Le Cunff et al. showed that the absence of structured classroom environments can exacerbate distraction, leading to frustration and disengagement~\cite{le_cunff2022supporting}. This issue is compounded in programming education, where tasks are often designed to require more sustained concentration and attention to detail than ADHD students are typically able to provide~\cite{zastudil2025neurodiversity}.

As educational platforms add video-based instruction, it is important to address the specific challenges and opportunities this medium presents for learners with ADHD. Previous work focusing on learners with ADHD in post-secondary education has shown that accommodations rarely support task management and attention regulation effectively~\cite{arnold2025beyond, ezeamii2025navigating}. Synchronizing verbal explanations with on-screen visuals reduces confusion and allows students to follow along with the instructor without extraneous cognitive effort~\cite{sweller2011cognitive}. Features such as adjustable playback speeds, rewind and replay options, and clearly segmented video content can provide students with greater control over the pace of instruction. However, prior work has shown that when students passively watch videos without pausing or reflecting, their learning outcomes are significantly diminished, and novice learners often do not know when or how to use these controls effectively, limiting their benefit even when such features are available~\cite{hundhausen-jvlc02, spanjers2011expertise}.

Our work focuses on individuals with ADHD, who often experience differences in regulation, executive function, and working memory~\cite{baddeley1992working, apa2025adhd} that can inhibit their ability to complete cognitively demanding programming tasks such as tracing programs or debugging. In remote learning environments, these learners must also independently regulate content pacing, manage outside distractions, and process unfamiliar concepts.

\subsection{Learning Interventions}

Video-based instruction is a crucial part of programming education in online educational settings. However, it is also a challenging instructional medium to work in, as videos should not try to replicate the in-person lecture experience. Instead, they should take advantage of their own unique affordances to impact students' education~\cite{begel2019outside}. A video's pace and structural organization plays a vital role in student comprehension. For example, Watkins et al. and Abdul-Rahman and du Boulay showed that shorter, well-structured instructional videos improve comprehension and reduce cognitive overload in programming education~\cite{watkins-2024, AbdulRahman2014}. Ljubojević et al. demonstrated that combining multimedia segmentation with pauses significantly reduced cognitive load and increased quality of experience in distance learning~\cite{ljubojevic2025improving}. Thorgeirsson et al. found that splitting video content into smaller, logically-organized blocks allowed learners to manage cognitive demands more effectively when completing programming tasks~\cite{thor-2024}.

Interactive videos demonstrating live coding activities have great promise to sustain students' attention while also giving them concrete coding examples they can try out on their own computers~\cite{manesh-icer2025}. Live coding places a higher cognitive demand on learners, however, who must simultaneously follow code creation and process verbal explanations~\cite{watkins-2024}. Manesh et al. confirmed that students often perceive live coding to be too fast-paced and caused them to experience increased mental workload when attempting to keep up with the instructional material~\cite{manesh-icer2025}. 

To mitigate such mental workload, we believe that the reduction of visual clutter through the elimination of extraneous interface elements can lower barriers to entry for students with varying cognitive needs. By simplifying the IDE layout in videos, we can improve \emph{Visibility} (ease with which relevant code can be found) and \emph{Diffuseness} (the number of distracting symbols a user needs to ignore to be able to process a single instruction)~\cite{rouly2014usability}. Green et al. highlights the importance of using ample white space, consistent formatting, and simple layouts to reduce distractions and enhance clarity~\cite{green1996usability, interactiondesignfoundation2016negative}. For example, breaking dense blocks of text into shorter paragraphs or bullet points and using visually clean diagrams could significantly improve comprehension. Similarly, in video-based instruction, eliminating extraneous elements on the screen and emphasizing critical steps through visual and auditory signals can prevent cognitive overload and keep students engaged.

\subsection{Interventions for Students with ADHD}

Zhu et al. highlight that video-sharing platforms pose unique challenges for individuals with ADHD, such as high information richness and less controllable content flow~\cite{zhu2025characterizing}. Some studies illustrate segmentation's potential as an effective strategy to address misaligned expectations for attention and promote more equitable learning outcomes for students with minority cognitive profiles, particularly those with ADHD. For example, Sechayk et al. introduced the ``SmartLearn'' and ``VeasyGuide'' tools, which leverage interventions like segmentation and highlighting to break down content and reduce cognitive load~\cite{sechayk2024smartlearn, sechayk2025veasyguide}.  Zhu et al. also created ``FocusView,'' a system enabling viewers with ADHD to customize multimodal elements such as background music and visual distractions to improve viewability~\cite{zhu2025focusview}. While FocusView addresses the diverse preferences of ADHD learners, its authors also noted that excessive customization options can themselves become a source of distraction. Additionally, previous studies have explored the capabilities of generative AI in supporting students with ADHD to design personalized learning management systems. However, findings show that even with higher perceived usability, human-created instruction remains vital to ensure effective education sessions~\cite{gunawardana2025enhancing, bernstein2025beyond, das2025towards}

While related work has explored interventions for students with ADHD, we follow Mayer’s segmenting principle which suggests breaking multimedia content into user-paced units generally improves learning outcomes~\cite{mayer2014multimedia}. However, Spanjers et al. indicate these benefits are often subject to an expertise reversal effect, primarily aiding those with lower prior knowledge or limited working memory capacity~\cite{spanjers2011expertise}. By targeting learners with no prior programming experience, our study specifically investigates segmentation at the point where the expertise reversal effect predicts the highest degree of intervention efficacy. 

We build upon the theoretical foundation and previous work by investigating logic-aware segmentation as a specialized compensatory mechanism, exploring how automated temporal buffers tied to discrete programming steps support the executive function constraints of ADHD learners.

\section{Study Design}

In our study, we applied an ad-hoc segmentation intervention onto our pre-recorded instructional videos on how to create a simple game in the platform Scratch. Pre-recorded video formats are particularly advantageous for learners with ADHD as they provide a stable, repeatable instructional stimulus that reduces the social and pacing pressures often found in live classroom environments~\cite{horlin2024normal}. As Horlin et al. highlight, the ability to control the pace of instruction allows neurodivergent students to bypass traditional classroom barriers and engage as ``normal'' students without the need for stigmatizing formal accommodations~\cite{horlin2024normal}.

\subsection{Instructional Videos}

\begin{figure}[tb]
    \centering
    \includegraphics[width=.45\textwidth]{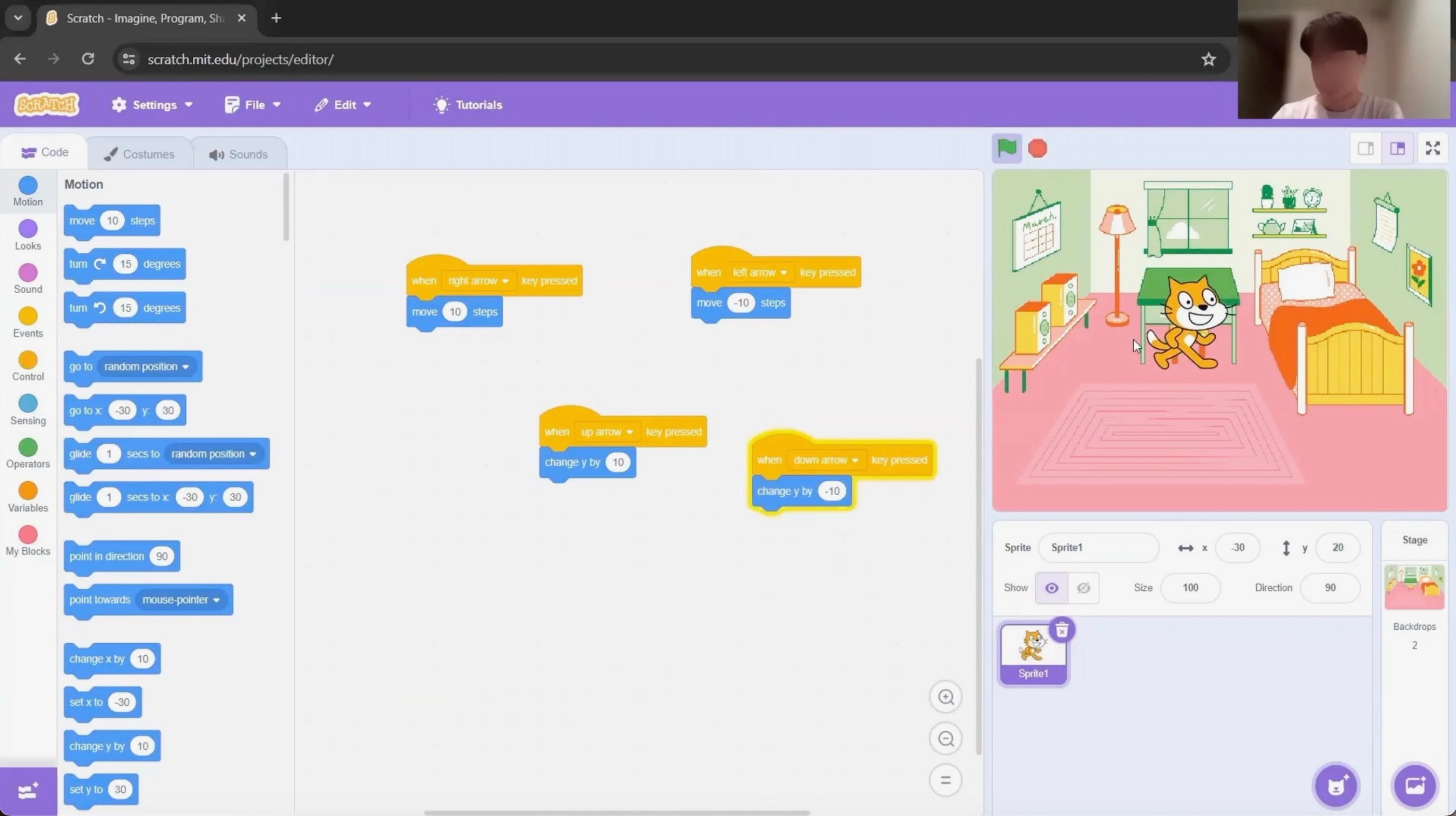}
    \Description{Screenshot of the Scratch programming interface showing movement control blocks, including commands to move a Cat sprite character up, down, left, and right when changed up, down, y, or x coordinated changed by 10, respectively.}
    \caption{Still from instructional video V2 used in the study, demonstrating movement controls in Scratch.}
    \label{fig:scratch-video}
\end{figure}

The primary instructional material consisted of six custom-made videos teaching the creation of a ``Cat and Mouse'' game in Scratch. We selected Scratch as the programming environment for this study because it is widely used in introductory programming education \cite{resnick2009scratch, scratchstats2026} for complete novices and minimizes syntax-related barriers, allowing us to focus on cognitive load and segmentation effects rather than programming language complexity. We intentionally avoided text-based languages to improve learnability for novices by favoring recognition over recall and reducing cognitive load by chunking computational patterns~\cite{bau2017learnable}. 

The videos depict screen recordings of an instructor demonstrating how to complete a task related to the game, with both verbal and auditory instruction. In this game, the participant controls a cat sprite to chase an automated mouse sprite; successful ``catches'' (collisions) increment a \textit{Score} variable by one point. Figure~\ref{fig:scratch-video} illustrates an example of Video V2, with blocks to control the cat's motion. 

Each video focused on a distinct set of programming skills:
\begin{itemize}
    \item \textbf{V1 (Background):} Navigating the interface to select and apply a new backdrop.
    \item \textbf{V2 (Movement):} Using event blocks (e.g., \textit{when key pressed}) and motion blocks to map arrow keys to X/Y coordinate changes. Repeat this for up, down, left, and right.
    \item \textbf{V3 (New Character):} Utilizing the sprite library to add a new character (the mouse).
    \item \textbf{V4 (Mouse Logic):} Implementing automated movement using \textit{forever} loops and the \textit{glide} block.
    \item \textbf{V5 (Variables):} Initializing a global \textit{Score} variable and setting its starting state.
    \item \textbf{V6 (Collision Logic):} Combining sensing blocks (\textit{touching}), conditionals (\textit{if-then}), and incrementing logic with a 1-second delay to prevent scoring loops.
\end{itemize}

The duration and granularity of these videos are detailed in Table~\ref{tab:video-stats}. To ensure a standardized classification of task difficulty (Easy, Medium, Hard), we examined the block and cognitive complexity ~\cite{sweller2011cognitive} from each video's programming skills. Easy tasks included the introduction and usage of 1 - 2 new buttons, medium tasks included the introduction and usage of 3 new buttons, and hard tasks included the introduction and usage of 4 or more new buttons.


\begin{table*}[t]
\centering
\caption{Instructional Video Properties and Segmentation Metrics}
\label{tab:video-stats}
\begin{tabular*}{\textwidth}{@{\extracolsep{\fill}}llcccc@{}}
\toprule
\textbf{Video} & \textbf{Video Skill / Concept} & \textbf{Original Length} & \textbf{Number of Segments} & \textbf{Difficulty} & \textbf{Length Post-Segments} \\ \midrule
V1 & Background \& Environment     & 00:34   & 8  & Easy   & 01:04 \\
V2 & Character Movement            & 01:49   & 22 & Medium & 03:17 \\
V3 & Adding a New Character (Sprite) & 00:25  & 5  & Easy   & 00:45 \\
V4 & Random Location Logic         & 00:48   & 10 & Medium & 01:28 \\
V5 & Variable Initialization       & 00:58   & 17 & Hard   & 02:04 \\
V6 & Collision \& Scoring Logic    & 01:08   & 14 & Hard   & 02:02 \\ \bottomrule
\end{tabular*}
\end{table*}


\subsection{Segmentation Intervention}

Our segmentation intervention is grounded in the \textit{segmenting principle}~\cite{mayer2014multimedia}: breaking complex information into learner-paced or system-paced chunks reduces extraneous cognitive load~\cite{merkt2018pauses}. While many design principles from the Cognitive Theory of Multimedia Learning (CTML) were originally developed using static media, Fyfield et al.~\cite{Fyfield2022} explained that segmenting remains one of the most empirically robust strategies specifically for improving learning gains in instructional videos. Segmented videos are significantly more effective than continuous formats, particularly when segments are aligned with logical learning activities~\cite{Fyfield2022}.

Following this evidence, we employed a \textit{logic-aware} segmentation approach to determine segment cuts: a pause was manually inserted immediately after the completion of a logical action (e.g., snapping a block into place), but before the explanation of the subsequent step. This ensures that each segment represents a complete semantic unit, allowing the learner to process the transient information before moving to the next task.

Crucially, our design utilizes \textit{system-defined} rather than student-initiated pausing~\cite{spanjers2011expertise}. This was an intentional design choice based on evidence that while student-initiated pausing often has negligible effects on learning, system-defined pauses can significantly improve outcomes for learners with lower working memory capacities~\cite{spanjers2011expertise}. By controlling the pause duration, our system ensures that interventions are applied consistently across all participants.

\subsection{Tasks}

To test our intervention, participants were given a set of six instructional videos in which each video taught a lesson towards the creation of a simple Scratch game. Each video was shown to participants in either segmented or non-segmented format, depending on the experimental condition. The segments were added manually with Final Cut Pro and agreed upon by multiple authors. After watching a video, participants then completed a task to assess their comprehension of the video's content. Each task was paired with a corresponding instructional video. There were six tasks in total, each matched to a specific instructional segment, covering progressively more complex elements of the Scratch game-building process: setting up a game environment, coding sprite movements, creating interactions, adding scoring mechanisms, and refining the game. We categorized tasks into easy, medium, or hard based on corresponding video complexity (see Table~\ref{tab:video-stats}), where all authors reviewed and agreed upon this classification. The specific sequential tasks and their assigned conditions are detailed in Table~\ref{tab:task_counterbalancing}.

\begin{table*}[t]
\centering
\caption{Task Breakdown, Complexity Tiers, and Counterbalancing Conditions}
\label{tab:task_counterbalancing}
\begin{tabular*}{\textwidth}{@{\extracolsep{\fill}}lllll@{}}
\toprule
\textbf{ID} & \textbf{Task Topic} & \textbf{Difficulty} & \textbf{Order A Condition} & \textbf{Order B Condition} \\ \midrule
T1          & Background Environment & Easy                & Non-segmented              & Segmented                  \\
T2          & Character Movement    & Medium              & Non-segmented              & Segmented                  \\
T3          & Second Character       & Easy                & Segmented                  & Non-segmented              \\
T4          & Random Location        & Medium              & Segmented                  & Non-segmented              \\
T5          & Score Initialization   & Hard                & Non-segmented              & Segmented                  \\
T6          & Collision Detection    & Hard                & Segmented                  & Non-segmented              \\ \bottomrule
\end{tabular*}
\end{table*}


\subsection{Pause Duration}

Before finalizing the intervention design, we conducted a pilot study with 13 participants to determine the optimal pause duration. We tested 2-second, 4-second, and 6-second intervals. Based on participant feedback and behavioral observations, 2-second pauses were perceived as too brief to allow for the mental rehearsal required for Scratch's spatial-logical mapping. Conversely, 6-second pauses led to noticeable drops in engagement and increased task-unrelated thoughts (mind-wandering). 

Following Merkt et al.~\cite{merkt2018pauses}, who suggest that learners with lower working memory capacity benefit most from structured pauses, we selected a 4-second duration. This interval was found to provide sufficient time for participants with ADHD to process the preceding instructional segment without introducing the fatigue or frustration associated with longer interruptions. This 4-second delay was inserted between every segment in the segmented condition, resulting in the total durations seen in Table~\ref{tab:video-stats}. We specifically matched the audiovisual content between conditions to ensure that any differences in performance were attributable solely to the temporal spacing of the segments.

\section{Hypotheses}

Prior literature on Cognitive Load Theory and Dual-Channel Processing led us to the hypothesis that a segmentation intervention could help reduce cognitive overload, particularly for individuals with ADHD. In addition, studies indicate that structured pauses may improve retention and engagement. However, the specific impact of pauses on individuals with ADHD in programming education remains underexplored. Within this context, we also consider medication as a mediating factor. 

\subsection{Behavioral Measures}

To empirically evaluate the cognitive effects, we operationalize cognitive load using two behavioral measures: \textit{errors} and \textit{hesitations}.

\textbf{Errors.} An \textit{error} is defined as any deviation from the demonstrated steps in the instructional video, including incorrect block placement, missing components, or unintended actions that alter the expected behavior of the Scratch program. 

\textbf{Hesitations}. A \textit{hesitation} is defined as a pause of 3 or more seconds during task performance or any verbal demonstration of confusion by the participant. This threshold was chosen based on prior research on cognitive load and working memory limitations, which suggests that prolonged pauses may indicate increased mental effort and difficulty in processing information~\cite{sweller2011cognitive, morrison-icer14}. Similar hesitation-based metrics have been used in prior studies assessing cognitive load in educational contexts~\cite{Betz2023}. The 3-second threshold was established through a consensus process among the five authors after analyzing participant behavior in the pilot study. We define verbal confusion as any audible utterance indicating a lack of task-clarity or a breakdown in self-regulation. This included explicit admissions of uncertainty (e.g., ``I don't know what to do next''), expressions of lost orientation (e.g., ``Wait, where am I?''), or redirected questions toward the researcher regarding instructions that had just been presented. By combining these qualitative markers with our objective 3-second temporal threshold, we were able to capture instances of cognitive overload that might not have resulted in a full behavioral pause but still indicated a significant disruption in the participant's mental model.

Each author independently reviewed pilot session recordings and identified points of noticeable hesitation. After discussion, we collectively agreed that hesitation-related pauses shorter than 3 seconds were common but did not consistently reflect cognitive overload, while hesitation-related pauses of 3 seconds or more were associated with visible confusion, uncertainty, or task difficulty. This definition aligns with Cognitive Load Theory (CLT), as increased cognitive demand can lead to observable delays in task execution when participants struggle to process and apply instructional content effectively. By setting a clear and empirically grounded threshold for hesitation, we aimed to ensure consistency and reliability in our analysis of cognitive load.

Utilizing these established definitions of errors and hesitations, we propose several detailed hypotheses to test the potential impact of our segmentation intervention, described next.

\subsection{Individual Hypotheses}\label{sec:hypo:ind}

\begin{enumerate}[label=\textbf{H\theenumi:},leftmargin=*,itemsep=1em]

\item \textbf{Task Difficulty Increases Errors and Hesitations}
\begin{itemize}[itemsep=0em]
    \item[\textbf{(a)}] Participants will make more errors on tasks that are harder compared to easier tasks.
    \item[\textbf{(b)}] Participants will make more hesitations on tasks that are harder compared to easier tasks.
\end{itemize}

\item \textbf{Segmentation Improves Performance for Participants with ADHD}
\begin{itemize}[itemsep=0em]
    \item[\textbf{(a)}] Participants with ADHD will make fewer errors after watching segmented videos compared to non-segmented videos.
    \item[\textbf{(b)}] Participants with ADHD will make fewer hesitations after watching segmented videos compared to non-segmented videos.
\end{itemize}

\item \textbf{Segmentation Benefits Participants with ADHD More than Participants without ADHD}
\begin{itemize}[itemsep=0em]
    \item[\textbf{(a)}] Participants with ADHD will show a larger reduction in errors due to segmentation than participants without ADHD.
    \item[\textbf{(b)}] Participants with ADHD will show a larger reduction in hesitations due to segmentation than participants without ADHD.
\end{itemize}

\item \textbf{Medication Impacts Performance for Participants with ADHD}
\begin{itemize}[itemsep=0em]
    \item[\textbf{(a)}] Medication impacts performance on the number of errors for participants with ADHD.
    \item[\textbf{(b)}] Medication impacts performance on the number of hesitations for participants with ADHD.
\end{itemize}
\end{enumerate}


\section{Experimental Method}

\begin{figure}[tb]
    \centering
    \includegraphics[width=.47\textwidth]{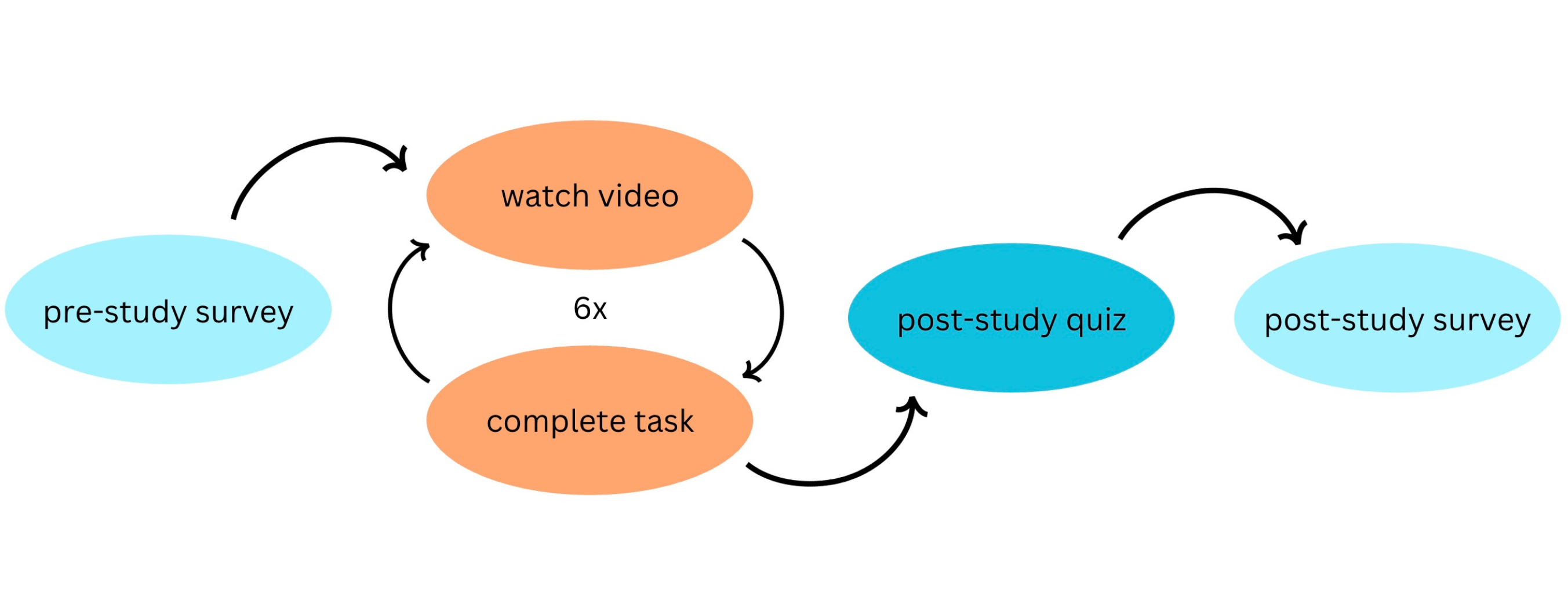}
    \Description{Study procedure including block for pre study survey, followed by an arrow to a repeating watch video then complete task (6x), then an arrow to post-study quiz and an arrow to post-study survey.}
    \caption{Study procedure outlining participant workflow through pre-study, task loop, and post-study assessments.}
    \label{fig:study-procedure}
\end{figure}

We conducted a within-subjects experiment to evaluate the impact of segmented instructional videos on individuals with ADHD in programming education. Study participants completed Scratch programming tasks under two counterbalanced conditions: unsegmented (i.e., continuous) videos and segmented videos with brief pauses after each instructional step. Our experimental design (shown in Figure~\ref{fig:study-procedure}) allows for a comprehensive analysis of how segmentation affects task performance and comprehension. 

The general study procedure included a pre-study survey followed by watching programming educational videos in Scratch (some of which included our segmentation intervention), completing a task corresponding to the video content, a post-study quiz, and a post-study survey.

\subsection{Participants}
The study consisted of 27 participants, divided into two groups: individuals with ADHD (n = 17) and individuals without ADHD (n = 10). Participants were required to be 18 years or older. None could have any programming experience to ensure that everyone started at the same baseline. This reduces variability due to prior knowledge and emphasizes the impact of segmentation on novice learners. Among participants with ADHD, 7 participants identified as female, 9 as male, and 1 as non-binary. Among participants without ADHD, 7 identified as female and 3 as male. Participant demographics, including whether participants with ADHD were taking ADHD medication at the time of their experimental session are shown in Table~\ref{tab:demographics}. 


Our ADHD group participants reflect the narrowing diagnostic gender gap observed in adulthood, where the ratio of affected individuals shifts from a male-heavy childhood distribution to approximately $1:1.6$ in adults~\cite{stibbe2020gender}. This representative distribution is particularly relevant to our analysis of processing speed and error rates; research suggests that adult females with ADHD may exhibit more unstable performance in working memory and inattentive symptoms compared to their male counterparts~\cite{stibbe2020gender}. By including a strong representation of female and non-binary participants, we ensure that the effectiveness of the 4-second pause intervention is evaluated against these documented cognitive variations rather than a male-centric hyperactive baseline.

Participants were recruited through flyers posted around nearby university campuses and email outreach by non-computer science faculty who shared the flyer with their students. Each participant received a \$25 USD Amazon gift card as compensation, which was emailed after completion of the one-hour in-person study session. IRB approval was obtained from our university for this study, ensuring compliance with ethical research standards.


\begin{table}[tb]
\centering
\caption{Participant Demographic Data ($N=27$)}
\label{tab:demographics}
\small 
\begin{tabular}{lllll}
\toprule
\textbf{ID} & \textbf{Gender} & \textbf{ADHD} & \textbf{Age Group} & \textbf{Medication} \\
\midrule
\rowcolor{gray!10} \multicolumn{5}{l}{\textit{ADHD Group}} \\
P1  & Male        & Yes & 25--34 & No \\
P2  & Female      & Yes & 18--24 & Yes \\
P3  & Male        & Yes & 45--54 & Yes \\
P4  & Non-binary  & Yes & 18--24 & No \\
P5  & Male        & Yes & 45--54 & No \\
P6  & Female      & Yes & 35--44 & Yes \\
P7  & Male        & Yes & 18--24 & Unknown \\
P8  & Male        & Yes & 18--24 & Unknown \\
P9  & Male        & Yes & 18--24 & Unknown \\
P10 & Female      & Yes & 18--24 & Yes \\
P11 & Male        & Yes & 25--34 & No \\
P12 & Female      & Yes & 25--34 & No \\
P13 & Female      & Yes & 18--24 & Yes \\
P14 & Male        & Yes & 18--24 & Yes \\
P15 & Male        & Yes & 45--54 & Yes \\
P16 & Female      & Yes & 18--24 & No \\
P17 & Female      & Yes & 18--24 & Yes \\
\midrule
\rowcolor{gray!10} \multicolumn{5}{l}{\textit{Control Group}} \\
P18 & Female      & No  & 18--24 & -- \\
P19 & Female      & No  & 18--24 & -- \\
P20 & Male        & No  & 18--24 & -- \\
P21 & Female      & No  & 18--24 & -- \\
P22 & Female      & No  & 55+    & -- \\
P23 & Female      & No  & 18--24 & -- \\
P24 & Female      & No  & 55+    & -- \\
P25 & Male        & No  & 25--34 & -- \\
P26 & Female      & No  & 25--34 & -- \\
P27 & Male        & No  & 18--24 & -- \\
\bottomrule
\end{tabular}
\end{table}

\subsection{Pre and Post Study Surveys}

Our study had two surveys: a pre-study survey before starting the tasks and a post-study survey after completing all tasks. The pre-study survey assessed participants' prior knowledge of programming, their confidence in using the platform, and their preferences for video-based instruction. The post-study survey collected feedback on the learning experiences of the participants, evaluating their understanding of the concepts and their perceptions of the segmented video instruction. It also gathered reflections on the effectiveness of the lesson in teaching Scratch concepts (when compared to the participants' baseline of no programming knowledge). 


\subsection{Tasks}

To account for potential \textit{order effects} in our six instructional video tasks, the segmentation order was partially counterbalanced across participants. Participants were randomly assigned to either Order A or Order B through an alternating assignment strategy, ensuring an equal distribution of conditions across both the ADHD and non-ADHD participants. This ensured that the order in which participants encountered the segmented and non-segmented tasks did not confound the results. The \textit{task order was also fixed}, with participants progressing through tasks in a predefined sequence, ensuring that they could build on their prior knowledge as they advanced through the tasks. 

Participants were not allowed to rewatch the instructional videos during task completion. The segmentation order was counterbalanced, while the task order was fixed to ensure participants built upon prior skills in a logical progression. See Table ~\ref{tab:task_counterbalancing} for details.

\subsection{Measures}

Several key measures were collected throughout the study to evaluate participant performance and the effects of video segmentation on task completion. Primary measures included the number of hesitations and errors. Additional measures recorded were task completion time, task correctness, and post-study quiz scores. 

All behavioral measures (errors and hesitations) were captured using a consensus coding approach~\cite{mcdonaldreliability}. By requiring agreement between two observers for every recorded data point, we established a shared `ground truth' that obviates the need for post-hoc inter-rater reliability (IRR) statistics. This collaborative method is particularly robust for identifying subtle behavioral cues, such as the 4-second hesitation threshold, as it minimizes individual subjectivity and ensures consistency across the dataset. Two authors counted the number of hesitations and errors for each participant in the same session. If discrepancies arose, they would discuss them and reach a consensus to ensure consistency and accuracy in qualitative coding. However, there were minimal disagreements in the final annotations.

\subsection{Quiz}

At the end of the study, par\-ti\-ci\-pants com\-plet\-ed a 12-question mul\-ti\-ple-choice quiz (described in Table~\ref{tab:quiz-summary}) to assess their conceptual understanding of the Scratch programming concepts taught during the study. This quiz ensured that par\-ti\-ci\-pants could apply what they learned rather than simply replicating actions from instructional videos. The full quiz is available in our replication package.\footnote{This study is designed to be fully replicable. A package containing all necessary materials, including data files, quiz and survey materials, and statistical analysis scripts is available on Zenodo at \url{https://doi.org/10.5281/zenodo.19701863}.}

\begin{table}[tb]
\centering
\caption{Summary of Post-Study Quiz Content}
\label{tab:quiz-summary}
\begin{tabular}{|l|p{5cm}|}
\hline
\textbf{Topic} & \textbf{Description / Example Items} \\
\hline
Scratch Blocks & Understanding movement blocks, loops (repeat/forever), conditionals (if/then), and coordinate changes (e.g., "change y by 10"). \\
\hline
Program Logic & Selecting the block that best fits inside a loop, reasoning about execution order, and comparing block functionality (True/False questions). \\
\hline
Variables & Understanding purpose and naming of variables (e.g., "Score" in the cat/mouse game). \\
\hline
Game Application & Applying block knowledge to simple game logic (e.g., movement of objects, scoring, and timing with "wait" blocks). \\
\hline
\end{tabular}
\end{table}

\subsection{Analysis}

To evaluate the effects of video segmentation on participant performance, we analyzed the number of errors and hesitations using Generalized Linear Mixed-Effects Models (GLMM). 
Because our data consists of non-negative count integers \textbf{N = 162 observations} collected through a within-subjects design, GLMMs allow us to account for the nested nature of repeated measures from the same participant by including random intercepts for each participant~\cite{bolker2009generalized}. 

We designed five fixed effect predictors based on the hypotheses we outlined in Section~\ref{sec:hypo:ind}:
$\beta_1(\text{Diff}_{ij})$ estimates the impact of task difficulty.
$\beta_1(\text{Seg}_{ij})$ estimates the impact of the segmentation intervention.
$\beta_1(\text{ADHD}_{j})$ estimates the baseline impact of ADHD status.
$\beta_1(\text{ADHD}_{j} \times \text{Seg}_{ij})$ estimates the interaction impact between ADHD status and the segmentation intervention.
$\beta_1(\text{Med}_{j})$ estimates the impact of medication status.
We use $i$ to represent the task index and $j$ to represent the participant index.

We fitted generalized linear mixed-effects models (GLMMs) to evaluate each hypothesis. For the error count outcomes, the models were defined using a Poisson distribution with a log-link function:
\begin{equation}
\ln(\mu_{ij}) = \beta_0 + f(\mathbf{X}_{ij}) + u_j
\end{equation}

\noindent where $\mu_{ij} = \mathbb{E}[y_{ij}]$ represents the expected error count for participant $j$ on task $i$, and $u_j \sim N(0, \sigma^2_u)$ is the random intercept accounting for repeated measures within participant $j$. The term $f(\mathbf{X}_{ij})$ denotes the linear combination of fixed predictors tailored to test each specific hypothesis:
\begin{align*}
    \mathbf{H_1}\text{:} \quad f(\mathbf{X}_{ij}) &= \beta_1(\text{Diff}_{ij}) \\
    \mathbf{H_2}\text{:} \quad f(\mathbf{X}_{ij}) &= \beta_1(\text{Seg}_{ij}) \\
    \mathbf{H_3}\text{:} \quad f(\mathbf{X}_{ij}) &= \beta_1(\text{Seg}_{ij}) + \beta_2(\text{ADHD}_{j}) + \beta_3(\text{Diff}_{ij}) \\
    &\quad + \beta_4(\text{ADHD}_j \times \text{Seg}_{ij}) \\
    \mathbf{H_4}\text{:} \quad f(\mathbf{X}_{ij}) &= \beta_1(\text{Seg}_{ij}) + \beta_2(\text{Med}_{j}) + \beta_3(\text{Diff}_{ij}) \\
    &\quad + \beta_4(\text{Med}_j \times \text{Seg}_{ij})
\end{align*}

For $H1$ we fit the GLMM with all tasks across Easy, Medium and Hard difficulty levels, but for $H3$ and $H4$  we only use tasks with Medium and Hard difficulties because we found that Easy tasks do not provide sufficient cognitive demand to differentiate the impact of the intervention, we discuss this in more detail in Section~\ref{sec:result:note}.

For the hesitation count outcomes, the structural linear predictors $f(\mathbf{X}_{ij})$ remained identical, but the models were instead fit using a Negative Binomial distribution ($y_{ij} \mid u_j \sim \text{NB}(\mu_{ij}, \theta)$) to account for observed overdispersion. This yields a conditional variance defined as:
\begin{equation}
\text{Var}(y_{ij} \mid u_j) = \mu_{ij} + \frac{\mu_{ij}^2}{\theta}
\end{equation}

\noindent where $\theta$ represents the estimated overdispersion parameter.





Given the sensitivity of $p$-values to sample size in interaction models, our interpretation of $H_3$ prioritized effect size magnitude and the direction of trends over strict null-hypothesis significance testing thresholds~\cite{SullivanFeinn2012, SanchezMeca2003}. Specifically, we interpreted Rate Ratios by relating them to standardized mean differences; following~\cite{Chen2010HowBI, hosmer2013applied}, Cohen’s $d$ benchmarks of 0.2 (small), 0.5 (medium), and 0.8 (large) were converted to equivalent ratio thresholds of 1.68, 3.47, and 6.71, respectively. 

Model outputs were summarized using estimated marginal means to interpret predicted counts, with pairwise contrasts and Tukey's HSD adjustments used to assess statistical significance. All analyses were run in R using the \texttt{lme4}, \texttt{glmmTMB}, and \texttt{emmeans} packages.

\subsection{Threats to Validity}

Regarding \textit{internal validity}, the fixed order of tasks (T1--T6) presents a potential maturation or order effect. We chose a fixed sequence to ensure a logical pedagogical progression, as the cumulative nature of the Scratch project required participants to master background and sprite creation before attempting complex scoring logic. We contend that counterbalancing task order was not feasible in this context, as skipping foundational steps would have rendered the subsequent instructional videos incomprehensible to novice learners. For \textit{construct validity}, we did not employ direct physiological or dual-task measures to quantify cognitive load. Kosch et al.~\cite{kosch2023survey} highlight a variety of metrics for measuring cognitive workload, including pupillometry, EEG, and secondary-task interference, but we prioritized non-invasive behavioral measures (errors and hesitations) to maintain ecological validity. Further, while we grouped participants by ADHD diagnosis, we did not directly measure individual working memory capacity using standardized assessments, such as an n-back test. Regarding \textit{external validity}, our sample of 27 individuals with no prior programming experience and our use of the Scratch block-coding environment may limit generalizability. However, the use of a Generalized Linear Mixed Model (GLMM) accounts for our data where each person had multiple observations for a total of $N = 162$ observations~\cite{bolker2009generalized}. 

Finally, while the 4-second pause duration was derived from pilot testing and established literature~\cite{merkt2018pauses}, it remains a fixed intervention. We acknowledge that individual differences in cognitive processing speed (even within neurodiverse groups) mean that a one-size-fits-all pause duration may not be optimal for every learner. To mitigate these threats, our methodological decisions are grounded in theory and previous work, and we have provided a full replication package including all survey instruments, anonymized data, and analysis scripts.

\section{Results}

The following section presents the statistical analysis of task performance across our four primary hypotheses. Analysis was conducted using Generalized Linear Mixed-Effects Models (GLMM) to account for the nested nature of the data (observations within participants). Errors were modeled using a Poisson distribution, while hesitations were modeled using a Negative Binomial distribution. Throughout the analysis, significance is evaluated through both $p$-values and standardized effect sizes (Cohen's $d$) to prioritize practical impact over sample-size-dependent metrics.

\subsection{H\texorpdfstring{$_1$}: Task Difficulty Increases Errors and Hesitations}

\begin{table}[tb]
\centering
\caption{GLMM Results for Task Difficulty on Performance Metrics ($H_1$)}
\label{tab:h1_difficulty}
\footnotesize 
\setlength{\tabcolsep}{2.5pt} 
\begin{tabular*}{\linewidth}{@{\extracolsep{\fill}}llcccc@{}}
\toprule
\textbf{Outcome} & \textbf{Predictor} & \textbf{Estimate ($\beta$)} & \textbf{Std. Error} & \textbf{$z$ value} & \textbf{$p$} \\ \midrule
\textbf{Errors} & (Intercept) [Easy] & -2.48 & 0.45 & -5.50 & < .001*** \\
($H_{1a}$) & Difficulty (Med) & 1.54 & 0.45 & 3.42 & < .001*** \\
& Difficulty (Hard) & 1.85 & 0.45 & 4.13 & < .001*** \\ \midrule
\textbf{Hesitations} & (Intercept) [Easy] & -2.24 & 0.43 & -5.18 & < .001*** \\
($H_{1b}$) & Difficulty (Med) & 2.24 & 0.46 & 4.89 & < .001*** \\
& Difficulty (Hard) & 2.67 & 0.46 & 5.87 & < .001*** \\ \bottomrule
\end{tabular*}
\begin{flushleft}
\scriptsize \textit{Note: $N=162$ observations. "Easy" serves as the reference level for all models.}
\end{flushleft}
\end{table}


\subsubsection{H1.a. Participants will make more errors on tasks that are harder compared to easier tasks}

The Poisson GLMM analysis revealed a significant main effect of task difficulty on error counts. Using Easy tasks as the baseline, participants committed significantly more errors on Medium tasks ($\beta = 1.54, p < .001$) and Hard tasks ($\beta = 1.85, p < .001$). Post-hoc pairwise comparisons using Tukey-adjusted estimated marginal means (EMMs) confirmed that error rates were significantly lower for Easy tasks compared to Medium ($p = .002$) and Hard tasks ($p < .001$). However, the increase in errors between Medium and Hard tasks did not reach statistical significance ($p = .472$), suggesting a potential plateau in error rates at higher difficulty levels.

\subsubsection{H1.b. Participants will make more hesitations on tasks that are harder compared to easier tasks. }

Consistent with the error analysis, a Negative Binomial GLMM showed that task difficulty significantly influenced hesitation frequency. Participants exhibited substantially more hesitations on both Medium tasks ($\beta = 2.24, p < .001$) and Hard tasks ($\beta = 2.67, p < .001$) relative to the Easy baseline. Pairwise comparisons confirmed that while Easy tasks were significantly more fluent than both higher difficulty levels ($p < .001$), the difference between Medium and Hard tasks was not statistically significant ($p = .178$). These findings support $H_{1b}$, indicating that increased task complexity leads to a significant reduction in behavioral fluency.

\subsection{Note on Easy (E) Tasks and Dataset Refinement}\label{sec:result:note}
Preliminary analysis of task performance following the validation of $H_1$ revealed a significant floor effect \cite{Everitt2010, Zhu2017} on Easy tasks across all participants. Regardless of ADHD status or segmentation condition, performance on these tasks remained consistently high, with low error means below 0.20 ($M_{\text{Control}} = 0.20$, $M_{\text{ADHD}} = 0.18$) and low hesitations showing a restricted range of [0--1]. Because these tasks did not provide sufficient cognitive demand to differentiate performance or measure the impact of the intervention~\cite{sweller2011cognitive}, they were excluded from subsequent analyses to avoid masking the effects observed under higher cognitive load. Following the removal of Easy tasks, the final dataset used for the testing of $H_2$, $H_3$, and $H_4$ consisted of $N = 104$ observations across the Medium and Hard difficulty levels.

\subsection{H\texorpdfstring{$_2$}: Segmentation Improves Performance for Participants with ADHD}

\subsubsection{H2.a. Participants with ADHD will make fewer errors after watching segmented videos compared to non-segmented videos.}

\begin{table}[tb]
\centering
\caption{GLMM Results for Segmentation Impact on ADHD Participants ($H_2$)}
\label{tab:h2_results_final}
\footnotesize 
\setlength{\tabcolsep}{2.5pt} 
\begin{tabular*}{\linewidth}{@{\extracolsep{\fill}}llcccc@{}}
\toprule
\textbf{Outcome} & \textbf{Predictor} & \textbf{Estimate ($\beta$)} & \textbf{Std. Error} & \textbf{$z$ value} & \textbf{$p$} \\ \midrule
\textbf{Errors} & (Intercept) & -0.47 & 0.37 & -1.30 & .194 \\
($H_{2a}$) & Segmentation & -2.05 & 0.53 & -3.85 & < .001*** \\ \midrule
\textbf{Hesitations} & (Intercept) & 0.70 & 0.19 & 3.64 & < .001*** \\
($H_{2b}$) & Segmentation & -1.54 & 0.29 & -5.37 & < .001*** \\ \bottomrule
\end{tabular*}
\begin{flushleft}
\scriptsize \textit{Note: $n=68$ observations ($17$ participants). Reference level is "Non-Segmented".}
\end{flushleft}
\end{table}


The model revealed a significant negative effect of video segmentation on error counts ($\beta = -2.05, SE = 0.53, z = -3.85, p < .001$). To interpret the practical impact, we calculated the Incident Rate Ratio (IRR) by exponentiating the coefficient ($e^{-2.05}$), resulting in an IRR of $0.128$. This indicates that when watching segmented videos, participants with ADHD produced approximately 87\% fewer errors compared to the non-segmented condition. The model showed no evidence of overdispersion, and the random effect for participants ($\sigma^2 = 0.97$) successfully captured individual differences in baseline error-proneness, providing strong support for $H_{2a}$.

\subsubsection{H2.b. Participants with ADHD will make fewer hesitations after watching segmented videos compared to non-segmented videos.}

The analysis for $H_{2b}$ revealed a highly significant negative effect of video segmentation on hesitation frequency ($\beta = -1.54, SE = 0.29, z = -5.37, p < .001$). By exponentiating the coefficient, we calculated an Incident Rate Ratio (IRR) of $0.214$, indicating that participants in the segmented condition exhibited approximately 79\% fewer hesitations than those in the non-segmented condition. 

The use of a Negative Binomial distribution was justified by the dispersion parameter ($\theta = 11.1$), which effectively modeled the variance. Furthermore, the participant-level random effect ($\sigma^2 = 0.24$) accounted for individual variability in hesitation patterns. These findings provide robust support for $H_{2b}$, suggesting that segmentation significantly reduces cognitive stalls during task execution.

\subsection{H\texorpdfstring{$_3$}: Segmentation Benefits Participants with ADHD More than Participants without ADHD}

\begin{table}[tb]
\centering
\caption{GLMM Interaction Results for Complex Tasks ($H_3$)}
\label{tab:h3_interaction}
\footnotesize 
\setlength{\tabcolsep}{2.5pt} 
\begin{tabular*}{\linewidth}{@{\extracolsep{\fill}}llcccc@{}}
\toprule
\textbf{Outcome} & \textbf{Predictor} & \textbf{Estimate ($\beta$)} & \textbf{Std. Error} & \textbf{$z$ value} & \textbf{$p$} \\ \midrule
\textbf{Errors}     & Segmentation (Seg) & -1.20 & 0.47 & -2.59 & .010** \\
($H_{3a}$)          & ADHD Status (Yes)  & -0.18 & 0.49 & -0.37 & .709   \\
                    & Difficulty (Hard)  & 0.34  & 0.27 & 1.29  & .197   \\
                    & Seg $\times$ ADHD  & -0.84 & 0.71 & -1.20 & .232   \\ \midrule
\textbf{Hesitations} & Segmentation (Seg) & -1.00 & 0.37 & -2.71 & .007** \\
($H_{3b}$)          & ADHD Status (Yes)  & 0.26  & 0.30 & 0.87  & .385   \\
                    & Difficulty (Hard)  & 0.43  & 0.20 & 2.11  & .035* \\
                    & Seg $\times$ ADHD  & -0.55 & 0.47 & -1.17 & .242   \\ \bottomrule
\end{tabular*}
\begin{flushleft}
\scriptsize \textit{Note: Reference levels are Non-Segmented, ADHD-No, and Medium Difficulty. $N=104$ observations.}
\end{flushleft}
\end{table}


\begin{table}[tb]
\centering
\caption{Magnitude of Segmentation Benefit by Group (Simple Effects)}
\label{tab:h3_simple_effects}
\footnotesize 
\setlength{\tabcolsep}{2.0pt} 
\begin{tabular*}{\linewidth}{@{\extracolsep{\fill}}llcccc@{}}
\toprule
\textbf{Outcome} & \textbf{Group} & \textbf{Rate Ratio} & \textbf{$p$} & \textbf{Cohen's $d$} & \textbf{\shortstack{Cohen's $d$ \\ Magnitude}} \\ \midrule
\textbf{Errors}      & Control (Non-ADHD) & 3.33 & .009** & 0.66 & Medium \\
                     & ADHD Group         & 7.75 & <.001*** & 0.85 & Large  \\ \midrule
\textbf{Hesitations} & Control (Non-ADHD) & 2.71 & .010* & 0.72 & Medium \\
                     & ADHD Group         & 4.70 & <.001*** & 1.14 & Large  \\ \bottomrule
\end{tabular*}
\begin{flushleft}
\scriptsize \textit{Note: Rate Ratio indicates the improvement factor under segmentation. Cohen's $d$ benchmarks: 0.2 = small, 0.5 = medium, 0.8 = large.}
\end{flushleft}
\end{table}


\subsubsection{H3.a. Participants with ADHD will show a larger reduction in errors due to segmentation than participants without ADHD}

The GLMM revealed a significant main effect of segmentation ($\beta = -1.20, p = .010$), confirming that the intervention reduced errors across the sample. While the interaction term ($\text{Seg} \times \text{ADHD}$) did not reach statistical significance ($\beta = -0.84, p = .232$), the substantive significance was evaluated by prioritizing effect size magnitude over the interaction $p$-value, as $p$-values are highly sensitive to sample size in complex interaction models~\cite{SullivanFeinn2012, SanchezMeca2003}.

Analysis of simple effects indicated that the reduction in errors for the ADHD group was characterized by a large magnitude (Ratio $= 7.75, d = 0.85, p < .001$), whereas the effect for the control group was notably smaller (Ratio $= 3.33, d = 0.66, p = .009$). Following the benchmarks established by Cohen, the intervention yielded a \textit{large effect} ($d > 0.8$) for participants with ADHD, providing a standardized benefit nearly 30\% stronger than that observed in the control group. These results suggest that segmentation effectively bridges the performance gap on complex tasks by providing a disproportionately robust scaffold for participants with ADHD.

\subsubsection{H3.b. Participants with ADHD will show a larger reduction in hesitations due to segmentation than participants without ADHD}

For hesitation frequency, the model identified a significant main effect of segmentation ($\beta = -1.00, p = .007$) and a significant effect of task difficulty ($\beta = 0.43, p = .035$), validating that high-difficulty tasks increased cognitive load. Although the interaction term was non-significant ($\beta = -0.55, p = .242$), group-specific analysis revealed a clear disparity in practical impact. 

The ADHD group exhibited an \textit{extremely large reduction} in hesitations (Ratio $= 4.70, d = 1.14, p < .001$), while the control group exhibited a medium-to-large effect (Ratio $= 2.71, d = 0.72, p = .010$). The substantial difference in both Rate Ratios and standardized effect sizes suggests that segmentation mitigates the executive function demands inherent in complex instructional materials more effectively for those with ADHD. While the study was underpowered to statistically distinguish the two groups through the interaction term, the magnitude of change provides practical support for the differential benefit hypothesized in $H_{3b}$.

\subsection{H\texorpdfstring{$_4$}: Influence of Medication Status on Intervention Efficacy}

To investigate whether medication status moderated the benefits of segmentation, we conducted a subgroup analysis on participants with ADHD ($n = 17$ participants, $68$ observations) focusing on complex tasks (Medium and Hard). Preliminary sensitivity analysis including participants as a third ``Unknown'' medication group revealed no significant differences or unique trends compared to the known groups; therefore, we removed them in the final analysis to maintain internal validity of the Medicated vs. Unmedicated comparison~\cite{Everitt2010}.

\begin{table}[tb]
\centering
\caption{GLMM Interaction Results for ADHD Subgroup ($H_4$)}
\label{tab:h4_interaction}
\footnotesize 
\setlength{\tabcolsep}{2.5pt} 
\begin{tabular*}{\linewidth}{@{\extracolsep{\fill}}llcccc@{}}
\toprule
\textbf{Outcome} & \textbf{Predictor} & \textbf{Estimate ($\beta$)} & \textbf{Std. Error} & \textbf{$z$ value} & \textbf{$p$} \\ \midrule
\textbf{Errors}      & Segmentation (Seg) & -1.54 & 0.64 & -2.42 & .016* \\
($H_{4a}$)           & Medicated (Yes)    & -0.31 & 0.64 & -0.48 & .632    \\
                     & Difficulty (Hard)  & -0.25 & 0.35 & -0.71 & .475    \\
                     & Seg $\times$ Med   & -1.10 & 1.21 & -0.90 & .366    \\ \midrule
\textbf{Hesitations} & Segmentation (Seg) & -2.03 & 0.48 & -4.19 & <.001*** \\
($H_{4b}$)           & Medicated (Yes)    & -0.71 & 0.34 & -2.07 & .039* \\
                     & Difficulty (Hard)  & 0.35  & 0.22 & 1.58  & .115    \\
                     & Seg $\times$ Med   & 0.89  & 0.62 & 1.43  & .153    \\ \bottomrule
\end{tabular*}
\begin{flushleft}
\scriptsize \textit{Note: Reference levels are Non-Segmented, Unmedicated, and Medium Difficulty. $N=68$ observations.}
\end{flushleft}
\end{table}


\begin{table}[tb]
\centering
\caption{Magnitude of Segmentation Benefit by Medication Status (Simple Effects)}
\label{tab:h4_simple_effects}
\footnotesize 
\setlength{\tabcolsep}{2.0pt} 
\begin{tabular*}{\linewidth}{@{\extracolsep{\fill}}llcccc@{}}
\toprule
\textbf{Outcome} & \textbf{Group} & \textbf{Rate Ratio} & \textbf{$p$} & \textbf{Cohen's $d$} & \textbf{\shortstack{Cohen's $d$ \\ Magnitude}} \\ \midrule
\textbf{Errors}      & Unmedicated & 5.00  & .016* & 0.79 & Medium/Large \\
                     & Medicated   & 14.00 & .011* & 1.09 & Large        \\ \midrule
\textbf{Hesitations} & Unmedicated & 7.58  & $<.001$*** & 1.52 & Large        \\
                     & Medicated   & 3.12  & .004** & 1.10 & Large        \\ \bottomrule
\end{tabular*}
\begin{flushleft}
\scriptsize \textit{Note: Rate Ratio indicates the factor by which performance improved under segmentation. Cohen's $d$ benchmarks: 0.2 = small, 0.5 = medium, 0.8 = large.}
\end{flushleft}
\end{table}


\subsubsection{H4.a: Influence of Medication Status on Error Reduction}

As shown in Table~\ref{tab:h4_interaction}, medication status was not a significant predictor of error counts among participants with ADHD ($\beta = -0.31, p = .632$). While the interaction term ($\text{Seg} \times \text{Med}$) did not reach statistical significance ($p = .366$), the simple effects in Table~\ref{tab:h4_simple_effects} demonstrate that the segmentation intervention provided a significant and large benefit to both subgroups. Medicated participants showed a numerically higher reduction in errors (Ratio $= 14.00, d = 1.09$) compared to unmedicated participants (Ratio $= 5.00, d = 0.79$). These results indicate that the cognitive scaffolding provided by segmented videos is robust and effective for participants with ADHD regardless of their pharmacological treatment status.

\subsubsection{H4.b: Influence of Medication Status on Hesitation Reduction}

For hesitation frequency, the model identified a significant main effect of medication status ($\beta = -0.71, p = .039$), confirming that medicated participants exhibited fewer hesitations overall. However, the segmentation intervention remained the primary driver of improved fluency ($p < .001$). 

Analysis of simple effects revealed a notable disparity in the magnitude of the benefit: unmedicated participants experienced an \textit{extremely large} reduction in hesitations (Ratio $= 7.58, d = 1.52$), while the benefit for medicated participants was comparatively lower (Ratio $= 3.12, d = 1.10$). Although the interaction term was non-significant ($p = .153$), the difference in effect size magnitudes suggests that segmentation acts as a powerful non-pharmacological scaffold. For students not utilizing medication, the intervention appears to provide the executive function support necessary to bridge the fluency gap, effectively bringing their performance closer to that of their medicated peers.

\subsection{Surveys and Post-Study Quiz}

To measure participants' self-efficacy, we collected their responses on pre- and post study surveys, where they rated their confidence on a 7-item Likert-scale survey. Across all tasks, including creating characters, using loops, and following video instructions, participants demonstrated an increase in self-efficacy. The aggregate average score rose from a pre-study mean of 2.51 ($SD = 1.05$) to a post-study mean of 4.34 ($SD = 0.59$). 

To assess retention of concepts learned, participants were given a post-study quiz. Overall, participants demonstrated strong knowledge of the programming material, achieving a \textbf{mean score of 9.67/12 ($SD = 2.05$)} and a \textbf{median of 10/12}. Individuals with ADHD reached a high level of proficiency with a mean of 9.62 ($SD = 2.18$) and individuals without ADHD reached a mean of 10.67 ($SD = 1.66$).

\subsection{Post-Study Interviews}
At the end of each session, participants responded to three  questions: (1) \textit{Did the lesson help you learn Scratch concepts? Please explain.} (2) \textit{What parts of the lesson helped you learn the most?} and (3) \textit{What reflections do you have on your experience in this study?}

Participants' answers revealed largely positive reflections on skill acquisition, detailing core structural understandings of loops, variables, and control structures. For example, P2 (ADHD) noted the lessons helped them \textit{``learn about looping motions,''} while P10 (ADHD) reported mastering variable and sprite creation. Conversely, the alienating effect of un-scaffolded multimedia instruction was highlighted by P12 (ADHD), who stated that \textit{``complicated instruction videos are really difficult to follow... I lost interest.''} 

When reflecting on impactful design features, participants noted that \textit{``pauses in the video''} (P21, Non-ADHD) allowed them to digest what the instructor was teaching ``so I could think and take in what he was attempting to teach me through the video'' (P5, ADHD), likely because they mitigated the intrinsic cognitive load of dual-channel processing. Further, participants noted that breaking complex sequences into discrete micro-steps was important. For example, P27 (Non-ADHD) said that in longer videos, ``I would have forgotten the earlier steps. I could remember where I was going, but not how to get there.'' We surmise that the micro-steps prevented working memory decay (similar to Watkins et al. and Manesh et al.~\cite{watkins-2024, manesh-icer2025}).

The feedback from our participants revealed a tension between perception and reality regarding our intervention's pauses. Three participants with ADHD noted that the silence during pauses disrupted their sustained attention. P2 (ADHD) reported finding it \textit{``harder to remember the steps with the pauses because then I would lose focus''}, while P6 (ADHD) noted that the initial pacing \textit{``broke my focus''} until they adapted to the format. P11 (ADHD) also reported focus difficulty during pauses. However, when we checked their actual performance, P2 improved on every task under the segmented condition, and P6 (ADHD) and P11 (ADHD) maintained stable performance across segmented and non-segmented conditions.

Participants described the cognitive limitations of asking them to repeat the steps they watched in a video. For example, P7 (ADHD) said these tasks were an implicit \textit{``test of memory.''} Several participants asked for tighter integration between instruction and execution. P22 (Non-ADHD) and P23 (Non-ADHD) asked for future intervention designs to actively pause instruction until a user completed the required action in the workspace in order to dynamically align the video with each individual's cognitive pacing.

\section{Discussion}

Our findings contribute to the understanding of how cognitive load and working memory capacity impact individuals with ADHD in asynchronous, video-based programming educational environments. In this section, we discuss the implications of our findings for ADHD-related theory, neurodiversity and disability, and educational practices more broadly.

\begin{enumerate}[label=\emph{H$_\theenumi$:},itemsep=0.5em,wide,labelindent=0pt]
\item \emph{Task Difficulty Increases Errors and Hesitations}
The significant increase in errors and hesitations as tasks moved from ``Easy'' to ``Medium'' difficulty confirms that programming tasks place a high demand on the executive functions of all learners. The stabilization in performance observed between ``Medium'' and ``Hard'' tasks (as evidenced by the overlapping marginal means in Table~\ref{tab:h3_simple_effects}) suggests a ``ceiling effect'' of cognitive load~\cite{salkind2010encyclopedia, sweller2011cognitive}. This indicates that learners likely reached their maximum processing capacity during the Medium difficulty phase. This alignment with Cognitive Load Theory (CLT) suggests that without instructional scaffolding, learners become overwhelmed once a task exceeds a certain threshold of intrinsic complexity, leading to the performance plateaus and behavioral hesitations observed~\cite{sweller2011cognitive, orru2019evolution}.

\item \emph{Segmentation Improves Performance for Participants with ADHD}
Our results indicate that the segmentation intervention significantly improved task performance for participants with ADHD, reducing errors by 87\% and hesitations by 79\%. While previous research has shown that pausing and ``chunking'' help learners generally~\cite{merkt2018pauses}, our study concretely demonstrates that this intervention is an effective, non-pharmacological, ad-hoc, video modification strategy in computing education. 4-second pauses allow for ``micro-restoration'' of the phonological loop and central executive components of working memory \cite{baddeley1992working}. This ``breathing room'' prevents the additive cognitive load of programming syntax from exceeding the learner's reduced working memory buffer~\cite{spanjers2011expertise, mayer2014multimedia}.

\item \emph{Segmentation as an Equalizing Intervention}
While the interaction term between segmentation and ADHD status did not reach statistical significance, the practical magnitude~\cite{SullivanFeinn2012, Chen2010HowBI} of the benefit was notably more pronounced for participants with ADHD. The Cohen’s $d$ for hesitations in the ADHD group ($d = 1.14$) suggests a large effect size that outperformed the benefit seen in the control group ($d = 0.72$). This suggests that the intervention acts as an equalizer, effectively closing the performance gap between neurodiverse and neurotypical learners. This outcome reinforces the core tenets of Universal Design for Learning (UDL)~\cite{rose2000universal} and the ``curb-cut effect''~\cite{reid2022curb}; while the segmented videos improved performance for all learners, they were essential for equity for students with ADHD. Our pedagogical strategy provides support without requiring learners to disclose their diagnosis or seek formal accommodations. This demonstrates that we do not need to identify who has ADHD in order to mitigate its associated barriers, as intentional design can bring the entire cohort to the same level of performance regardless of individual neurocognitive profiles.

\item \emph{Medication and Structural vs. Behavioral Constraints}
The intervention was highly effective regardless of pharmacological status, which suggests that segmentation targets structural working memory constraints~\cite{baddeley1992working} rather than behavioral attention alone. While medication may improve a learner's ability to ``stay on task'' (behavioral), the 4-second pause physically reduces the amount of information the brain must hold at one time (structural). For unmedicated participants, who experienced the largest magnitude of benefit ($d = 1.52$), the intervention provides an externalized executive function scaffold, performing the ``filtering'' and ``sorting''~\cite{diamond2013executive}.
\end{enumerate}

\subsection{Theoretical Implications Related to ADHD}

Our findings conform with predictions made by Cognitive Load Theory (CLT), Dual-Channel Processing, and Working Memory Theory. According to CLT, learning is most effective when extraneous cognitive load (the effort required to process instruction) is minimized, and intrinsic load is appropriately scaffolded~\cite{sweller2011cognitive}. The significant improvements observed during medium and hard tasks suggest that segmentation is most effective when learners are cognitively challenged. This aligns with CLT's premise that instructional interventions yield greater benefit when intrinsic load increases. The reduction in hesitations further suggests that video pauses gave learners with ADHD sufficient time to rehearse and encode the instructor's verbal commands, alleviating demands on the phonological loop, which supports verbal processing in working memory~\cite{baddeley2000episodic}. This interpretation aligns with cognitive models emphasizing the temporal limits of working memory~\cite{baddeley1992working}. 

When considering the effect of stimulant medication, if it primarily supports sustained attention without expanding working memory capacity~\cite{advokat2010cognitive}, then pharmacological and instructional interventions likely operate through different mechanisms. In this context, segmentation functions not to compensate for shorter attention spans, but as a design strategy that reshapes the pacing of instructional content. By introducing a brief pause, the video structure provides learners with the temporal space needed to rehearse and encode each verbal command into long-term memory before it decays from the phonological store~\cite{baddeley1992working}. The four second pauses likely served as a temporal buffer which prevented a ``split attention'' effect (inherent in dual-channel processing) and allowed participants to offload visual-spatial information from the Scratch interface before the next verbal instruction was given~\cite{vass-2011}. By applying CLT, dual-channel processing, and working memory theory to this context, our study fills a gap in the existing literature and points to the potential of more adaptive, cognitively aligned interventions in computing education for individuals with varying cognitive needs.

\subsection{Implications for ADHD and Neurodiversity}

While our findings suggest that pacing of instruction for programming education could impact learners with ADHD, segmentation is only one of many potential post-hoc video adjustments that may alleviate cognitive overload. Interventions like Kasatskii et al.'s \cite{kasatskii2023effect} ``Low Perceptual Load Mode,'' removed extraneous visual ``noise'' from the user interface of an IDE to reduce distraction. One could also  highlight salient parts of a UI during an instructional task to make it easier to follow a clear visual path. 

While our study focused on individuals with ADHD, temporal segmentation could be applied as a ``post-hoc'' modification of pre-recorded programming educational videos for a wider range of neurodivergent cognitive profiles, such as reduced phonological processing efficiency, detail-focused processing, or slower cognitive processing speeds. For example, individuals with reduced phonological processing efficiency (e.g., dyslexia) often face higher cognitive load when decoding text-based syntax~\cite{powell2004dyslexia}.  In programming videos, reading on-screen text may function as a secondary task required for learners to identify the object referenced by the instructor. However, because this decoding effort is incidental to the learning objective, it introduces extraneous cognitive load. Highlighting the referenced on-screen element can reduce this burden and improve accessibility. For people who process sensory input in a detail-focused manner, commonly observed in autistic people, auditory and visual clutter can be particularly distracting~\cite{israel2020descriptive, ferreiraderaposo2024hci}. Reducing such clutter may help them focus on the more salient aspects of instruction. Furthermore, for learners with slower cognitive processing speeds, adding persistent captions that track the instructor's progress through the video can shift some processing demands from the auditory to the visual channel by providing a consistent visual anchor.

\subsection{Implications for Other Disabilities}

Beyond neurodivergence, temporal scaffolding through segmentation could extend to individuals with physical disabilities. One could extend Li et al.'s work on non-visual cooking instruction for people with visual impairments with automated pauses to account for additional time required to physically locate tools~\cite{li2021nonvisual}. Similarly, Sechayk et al. introduce a system which addresses visual search challenges faced by low vision learners when instructors perform subtle actions such as pointing or sketching without verbal descriptions through highlighting and magnifying important visual cues~\cite{sechayk2025veasyguide}. An intervention could add strategically placed pauses to videos with highlights and magnification to transition learner's gaze and allow time to stabilize their visual focus on the highlighted and magnified areas.

\subsection{Implications for Adaptive Interventions}

Future interventions could transition from static pauses to closed-loop pacing, where videos auto-pause after a command and resume only when the system detected that the learner completed the action in their own workspace. This could be augmented with visuals that reduce distraction by dynamically highlighting referenced UI elements, magnifying subtle instructor actions, and reducing unnecessary visual clutter~\cite{sechayk2025veasyguide, kasatskii2023effect}. This feature could be automated with an AI-based segmentation tool trained on public datasets (such as YouTube~\cite{youtube} or Khan Academy~\cite{khanacademy2025} tutorials) to detect appropriate pause boundaries by analyzing instructor speech, cursor movements, and code execution. Finally, to address the way some learners objectively benefit from pauses yet subjectively feel a loss of focus, future instantiations should increase user control by supporting customization of pause durations or even complete deactivation of pauses.



\section{Conclusion \& Future Work}

This study contributes to the growing body of research on inclusive educational technologies, demonstrating the effectiveness of segmented instructional videos for individuals with ADHD. By breaking complex tasks into smaller, more digestible steps, segmentation reduces cognitive load and eases the burden on working memory, lowering error rates and supporting task completion. Our findings emphasize the need for adaptive learning environments that accommodate the diverse cognitive abilities of students. 


While the implementation of the intervention for this study required manually editing instructional videos to insert pauses, future work could explore the use of LLMs and GenAI to automate this process. An automated pipeline could identify segments where the instructor rapidly issues commands to the learner, and insert pauses after each. AI that could recognize learner actions in their own programming environment could enable automatically restarting paused video instructions whenever the appropriate actions have been successfully completed. This could be especially helpful for more complex programming concepts such as algorithm complexity, data structures, and object-oriented programming.

Overall, this study provided empirical evidence through a study with 27 individuals with and without ADHD to show that simple, post-hoc segmentation of instructional videos can reduce extraneous cognitive load and support individuals with ADHD to retain unfamiliar programming concepts. Our segmentation intervention assists in reducing the environmental barriers imposed on students with ADHD in traditional learning settings. 
We hope this work will inspire future research to move beyond general accessibility and toward the development of neuro-inclusive systems that treat cognitive support as a fundamental design imperative.




\bibliographystyle{ACM-Reference-Format}
\bibliography{references, software}


\end{document}